# A New Strained-Silicon Channel Trench-gate Power MOSFET: Design and Analysis

Raghvendra S. Saxena and M. Jagadesh Kumar, *Senior Member, IEEE*

**Abstract:** In this paper, we propose a new trench power MOSFET with strained Si channel that provides lower on resistance than the conventional trench MOSFET. Using a 20% Ge mole fraction in the $Si_{1-x}Ge_x$ body with a compositionally graded $Si_{1-x}Ge_x$ buffer in the drift region enables us to create strain in the channel along with graded strain in the accumulation region. As a result, the proposed structure exhibits 40% enhancement in current drivability, 28% reduction in the on-resistance and 72% improvement in peak transconductance at the cost of only 12% reduction in the breakdown voltage when compared to the conventional trench gate MOSFET. Furthermore, the graded strained accumulation region supports the confinement of carriers near the trench sidewalls improving the field distribution in the mesa structure useful for a better damage immunity during inductive switching.

I. INTRODUCTION

A trench gate MOSFET [1-15] is the most preferred power device for medium to low voltage power applications. These are used extensively in control switching, DC-DC converters, automotive electronics, microprocessor power supplies etc. In all these applications, low on-state resistance is the prime requirement to reduce the conduction power loss and forward voltage drop. Higher drive current, low gate-to-drain capacitance, high transconductance, high breakdown voltage and inductive switching capability are the other requirements in various applications of power MOSFETs [16-20]. Different techniques have been proposed for reducing the on-state resistance and improving other performance parameters [5-13, 19-22]. Out of various components of the total resistance, the channel resistance is the biggest resistance contributor and needs to be suppressed without significantly affecting the other performance parameters. Among the techniques of reducing channel resistance, the use of $Si_{1-x}Ge_x$ channel has been reported to give up to 10% improvement in the on-resistance [8]. The strained Si channel is also used to significantly improve current drivability and transconductance in lateral power MOSFETs [20, 21]. However, the same is not feasible in a conventional trench structure as the formation of strained Si channel results in the elimination of the accumulation region, reducing its damage immunity for inductive load switching, which is also an essential requirement in some applications [9]. Therefore, the main objective of this paper is to propose an improved trench gate MOSFET structure that offers lower on-resistance by allowing the formation of strained Si channel and efficient confinement of the carriers near the trench side-walls in the accumulation layer needed for improved inductive switching capability.

In this paper, we present the structure of the proposed device with its fabrication feasibility. Using 2D numerical simulations performed with ATLAS device simulator [23], we present an extensive analysis of the proposed device in contrast with the conventional device, showing that the proposed device exhibits improved drive current and transconductance and reduced on state resistance as compared to the conventional device.

## II. DEVICE STRUCTURE AND PROPOSED FABRICATION PROCEDURE

Fig. 1 shows the cross-sectional view of the proposed device structure, termed here as SCT (Strained-Si Channel Trench) MOSFET. As apparent from the figure, the SCT-MOSFET uses P-type $Si_{0.8}Ge_{0.2}$ in the body and a compositionally graded N-type $Si_{1-x}Ge_x$ buffer layer (x = 0.0 at the Si drift region side and x = 0.20 at the body side) in the drift region. The $Si_{1-x}Ge_x$ buffer layer in the drift region serves three purposes. First, it allows the growth of defect free $Si_{0.8}Ge_{0.2}$ body that is required for the strained Si channel formation. Secondly, it causes graded strain in Si accumulation layer that results in smoothing of the conduction band discontinuity between strained Si channel and Si-drift region, eliminating the problem of carrier transport due to conduction band discontinuity between these two regions. The graded strain also provides carrier confinement in the accumulation region, resulting in the electric field relaxation in the mesa structure, which is good for inductive load switching [9].

The proposed fabrication procedure of the SCT structure is similar to the methods used to fabricate the conventional trench gate MOSFET till some initial steps [4, 6, 8, 12, 14, 15]. We start with the $N^+$ ($N_D = 1 \times 10^{19}$ cm$^{-3}$) wafer and grow a 2.5 μm thick $N^-$ Si epilayer ($N_D = 1 \times 10^{16}$ cm$^{-3}$). This layer forms the drift region of the device. Over this layer we grow a 0.5-μm thick N-type $Si_{1-x}Ge_x$ buffer layer ($N_D = 1 \times 10^{19}$ cm$^{-3}$) with gradually changing the x-composition from 0 to 20%. After this a 0.4 μm thick P-type $Si_{0.8}Ge_{0.2}$ body ($N_A = 5 \times 10^{17}$ cm$^{-3}$) and 0.1 μm thick $N^+$ $Si_{0.8}Ge_{0.2}$ source ($N_D = 1\times10^{19}$ cm$^{-3}$) are grown. Now, a 1.0 µm wide and 1.2 µm deep trench is opened as shown in Fig. 2(a). A 20 nm thin N-type ($N_D = 1\times10^{16}$ cm$^{-3}$) Si epi-layer is grown in the trench. The part of the Si epilayer touching the $Si_{1-x}Ge_x$ layer becomes strained. The trench is then filled with the deposited sacrificial oxide. The 0.5 µm deep trench is again opened with the same trench mask that removes the N-strained layer also from the side walls. After that, we selectively grow the P-strained Si epilayer that forms the channel of the device, as shown in Fig. 2(b). A 50 nm thick gate-oxide layer is then deposited. We recommend here an initial thermal growth of oxide up to a few angstroms and then the oxide deposition. This approach results in good interface without consuming much Si. The trench is then filled with N$^+$ poly Si as a gate material, as shown in Fig. 2(d). The formation of channel is self-aligned and therefore it makes the process immune to the variations in the depth of reopened trench. According to our simulations a 10% variation in the depth of trench results in only less than 0.6% variations in various parameters. Finally, the source, drain and gate contacts are taken and the device structure becomes like the one shown in Fig. 1.

### III. RESULTS AND DISCUSSION

In ATLAS device simulator we have created the SCT device with various layers and doping concentrations as discussed above. We have first created the graded $Si_{1-x}Ge_x$ buffer and the graded strained Si layers in the drift region by using ten $Si_{1-x}Ge_x$ layers of 50 nm thickness with different x composition values changing from x=0.0 at the bottom of the layer to x=0.2 at the top of the layer in 10 uniform steps. On the trench side of these $Si_{1-x}Ge_x$ layers, we created ten corresponding 20 nm wide layers of strained Si. For the realization of strained Si in the simulator, we have modified the energy band structure (electron affinity and energy band gap) and low field mobility in each of these layers according to their respective strain as done in previous works [24-27]. Similarly, we have created the P-type $Si_{0.8}Ge_{0.2}$ body and corresponding strained Si layer in the channel region. The SCT MOSFET device has been simulated and analyzed for its energy band diagram, current-voltage characteristics and breakdown performance. Since the contact resistance is a negligible contributor (usually less than 5 %) of the total on-resistance, we have assumed the contact resistance to be negligible for both the devices. To the best of our knowledge, we do not know of a model of impact of temperature on the energy bands of strained silicon that can be used in device simulation for studying the thermal issues. Therefore, we have not carried out any studies on the thermal effects.

The simulation results as compared with those of the conventional device having similar geometry and doping parameters are discussed below.

A. *Effect of Energy Band Modifications*

The strain in the channel and in the accumulation region causes modifications in the energy band structure. The simulated energy band structure is calculated along the cut lines A, B (both in transverse direction to the current flow) and C (along the current flow) as marked in Fig. 1. Fig. 3(a) shows the comparison of energy band structures of the proposed SCT device and the conventional device in the channel along cut line A for typical bias condition of $V_{GS}$ = 5 V and $V_{DS}$ = 0.1 V. The negative valance band offset causes the Fermi level to shift towards the conduction band contributing more electrons in the channel for the same gate bias, resulting in a threshold voltage shift. Our simulations indicate a shift in threshold voltage from 2.1 V in conventional device to 1.5 V in the proposed SCT device. The use of graded strained Si in accumulation region removes the abruptness in the conduction band discontinuity from the carrier transport path between the strained channel and the unstrained drift region, as illustrated in Fig. 3(b), showing the conduction band energy (plotted along the cut line C) for the proposed device along with that of the one having no strain in the accumulation region. It is also evident that by using graded strain Si layer, the potential barrier of about 0.11 eV due to abrupt discontinuity has been reduced to a gradually increasing barrier of 0.06 eV in the proposed device supporting the smooth transition of the energy bands and hence the carrier transport.

The conduction band discontinuity due to hetero-structure formation from channel to body region helps the carrier confinement in the channel as shown in Fig. 4(a) that shows the comparison of carrier concentration profiles for SCT and conventional devices for the same gate overdrive voltage of 5 V, along the cut line A. The carriers are also confined in the gradually strained accumulation region and this confinement reduces as we go deeper in y direction. Fig. 4(b) shows the carrier profile typically at 0.3 µm deep cut line B from the body in the x direction to show the carrier confinement in the accumulation region.

B. *Current Voltage Characteristics*

The output characteristics ($I_{DS}$-$V_{DS}$) for the SCT device and the conventional device are shown in Fig. 5(a), depicting the higher drive current in SCT device as compared to the conventional device for all bias conditions. The transfer characteristics ($I_{DS}$-$V_{GS}$) for these devices are shown in Fig. 5(b) for small $V_{DS}$ (0.1 V to 0.5 V) which is the usual operating condition of an on state power MOSFET [16, 28]. The on state resistance of the device is the ratio of applied $V_{DS}$ to the resulting $I_{DS}$ in the linear region of operation and it varies with the applied $V_{GS}$ [16]. The on state resistance evaluated at $V_{DS}$ = 1 V, as a function of gate voltage for the SCT device in contrast to the conventional device is shown in Fig. 5(c). As expected, the SCT device shows lower on resistance as compared to the conventional device. The figure also shows the percentage improvement in the on resistance of the device. As the gate voltage increases, the high transverse electric field tends to reduce the mobility in the channel for both the SCT and conventional devices. Therefore, beyond a certain gate voltage, the strain induced mobility enhancement factor (that is responsible for current enhancement in SCT device) reduces resulting in a lesser improvement in the drive current and on-state resistance as compared to the conventional device. However, at a gate voltage of 5 V, we observe from Fig. 5(c) that the reduction in on-resistance is approximately 28 % as compared to the conventional trench MOSFET. This is an acceptable improvement since the on-resistance of MOSFETs approximately depends on the $2.5^{th}$ power of breakdown voltage reduction, in general.

Furthermore, the proposed SCT device shows an excellent peak transconductance ($g_m$). This occurs due to the potential well formation in the channel of SCT device. The resulting carrier confinement causes more number of carriers to respond to the small signal voltage applied at the gate as compared to the conventional device. As a result, we get larger $g_m$ at lower gate overdrive voltages and about 72% improvement in peak $g_m$ in SCT device as compared to the conventional device as shown in Fig. 5(d), making it better for amplification purpose.

C. *Drain Breakdown Voltage*

At breakdown condition, a significant current starts flowing between drain and source by avalanche multiplication process [16]. Practically, the breakdown voltage is reported as the drain to source voltage at which $I_{DS}$ crosses a certain limit in the off condition i.e., with the gate tied to the source. The lower energy bandgap in $Si_{1-x}Ge_x$ as compared to Si results in higher avalanche multiplication factor and causes a reduction in the breakdown voltage in SCT device. We found a 12% reduction in the breakdown voltage of SCT device, compared to the conventional device, as shown in Fig. 6. Here, we have selected breakdown limit of drain current to be 10 pA/μm.

Thus, in SCT device we get better performance as compared to the conventional device in terms of large currents, low on state resistance and high transconductance with a small degradation in breakdown voltage.

## IV. CONCLUSIONS

Using 2-D numerical simulations, we have demonstrated that strain can be introduced in the channel of a trench gate power MOSFET by using $Si_{1-x}Ge_x$ body leading to improvements in the device performance. The use of 20% Ge mole fraction in the body with a graded $Si_{1-x}Ge_x$ composition in the drift region results in the strained Si channel and graded strained accumulation region giving quantifiable bench-marks of drive current improvement of 40%, the on state resistance reduction of 28% and the peak transconductance improvement of 72% as compared to the conventional trench gate MOSFET device. The demonstrated improvement in the performance of trench gate power MOSFET using strained silicon channel is expected to provide the incentive for experimental verification [29].

FIGURE CAPTIONS

Fig. 1: The cross-sectional view of the proposed strained Si channel trench (SCT) MOSFET device.

Fig. 2: Proposed fabrication process steps for SCT device.

Fig. 3: The energy band diagram of SCT and the conventional devices (a) The conduction band and valence band edges along cut line A showing Fermi level shifting, (b) Conduction band edge of proposed SCT device along cut line C in comparison with the device having abrupt transition between channel and drift region.

Fig. 4: Carrier concentration profiles of SCT and conventional devices (a) along cut line A, (b) along cut line B, showing the confinement of carriers in SCT device as compared with the conventional device.

Fig 5: The device terminal characteristics comparing the conventional and the SCT devices (a) Output characteristics, (b) Transfer characteristics, (c) On-state resistance as function of gate voltage ($1^{st}$ Y-axis) with $V_{DS}$ = 1.0 V and percentage reduction in on state resistance ($2^{nd}$ Y-axis) as compared to the conventional device, (d) Transconductance of the SCT and the conventional devices as functions of gate voltage.

Fig. 6: The breakdown performance of the SCT and the conventional devices for $V_{GS}$ = 0 V.

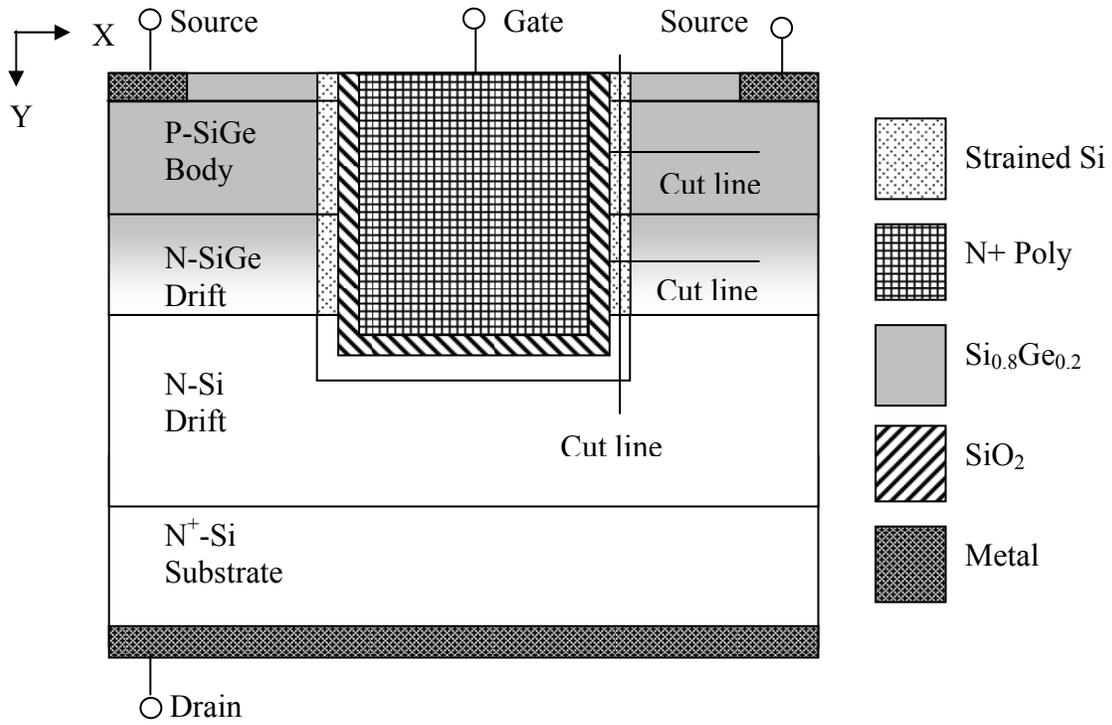

Fig. 1

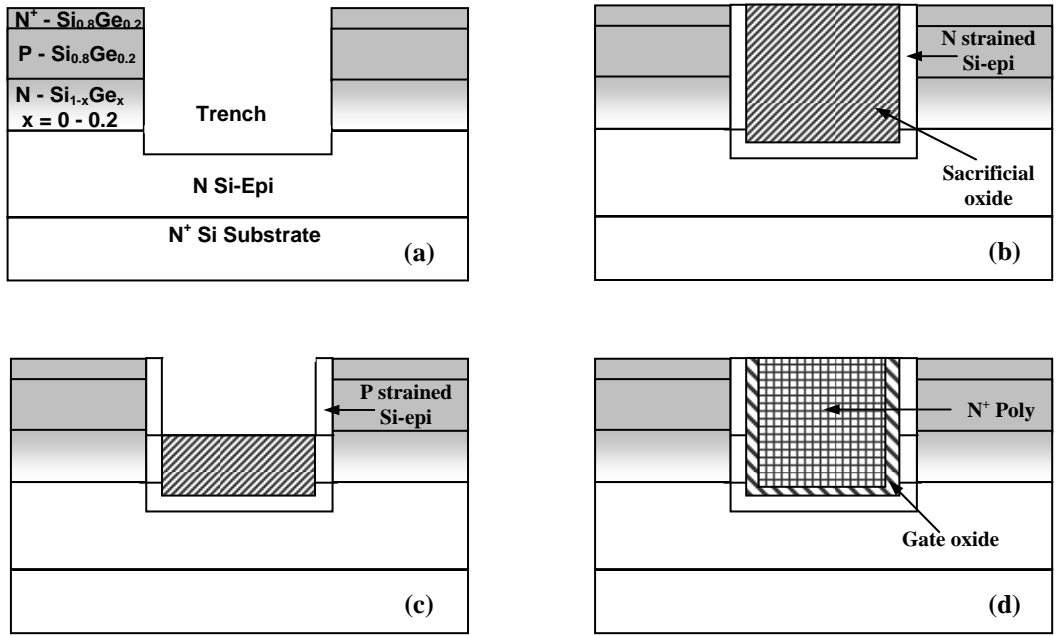

Fig. 2

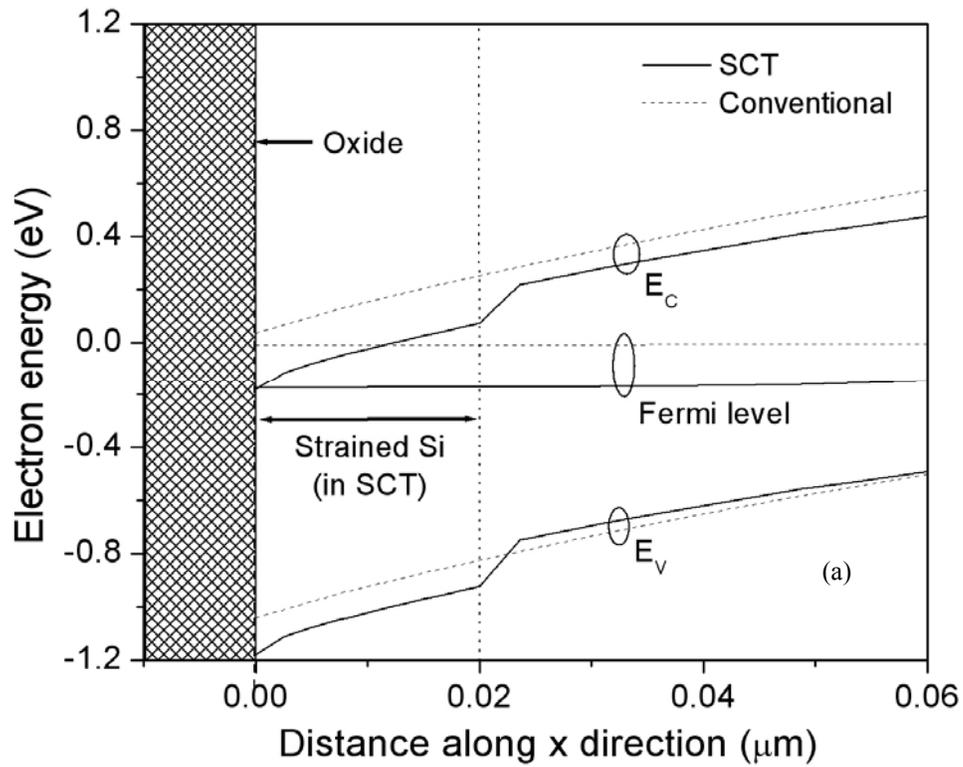

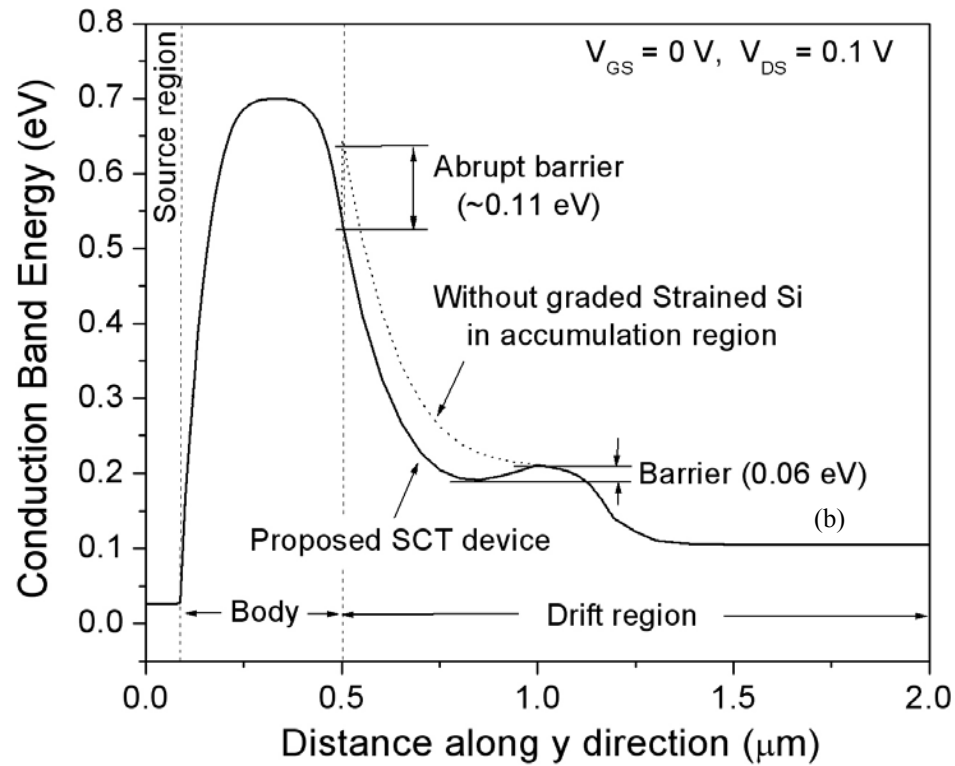

Fig. 3

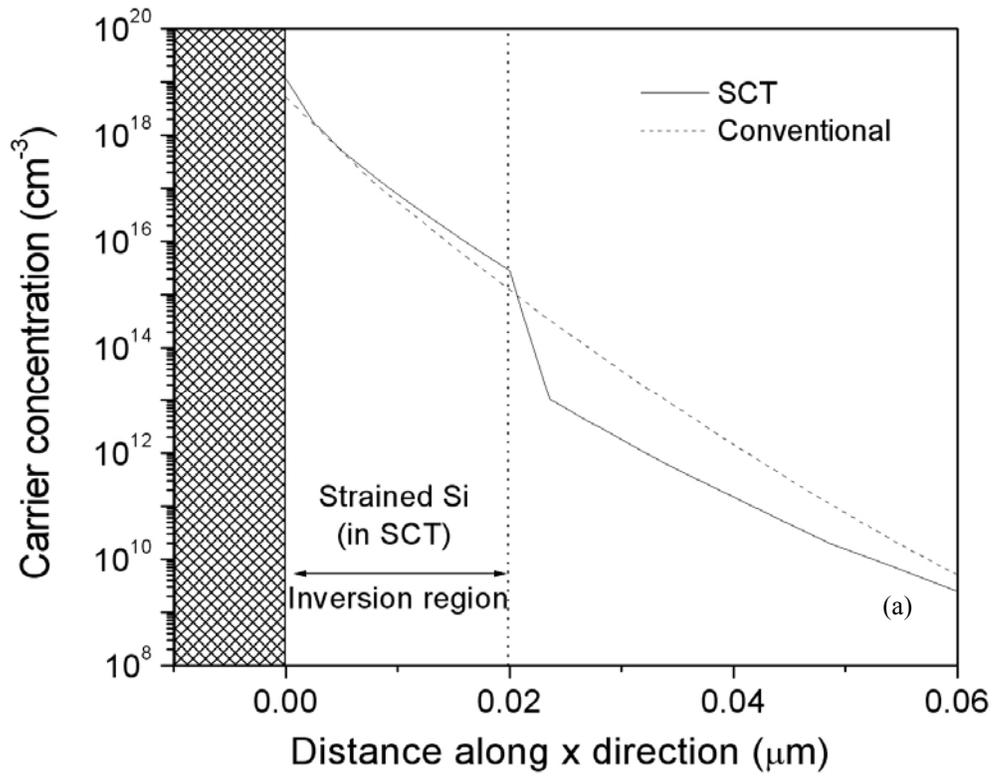

(a)

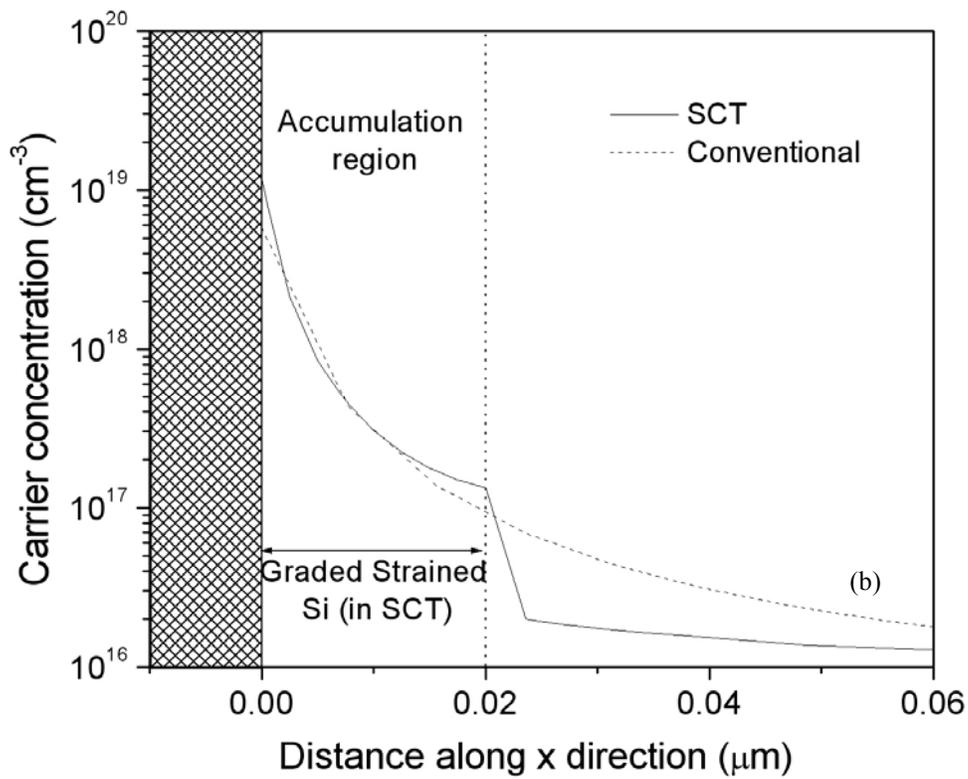

(b)

Fig. 4

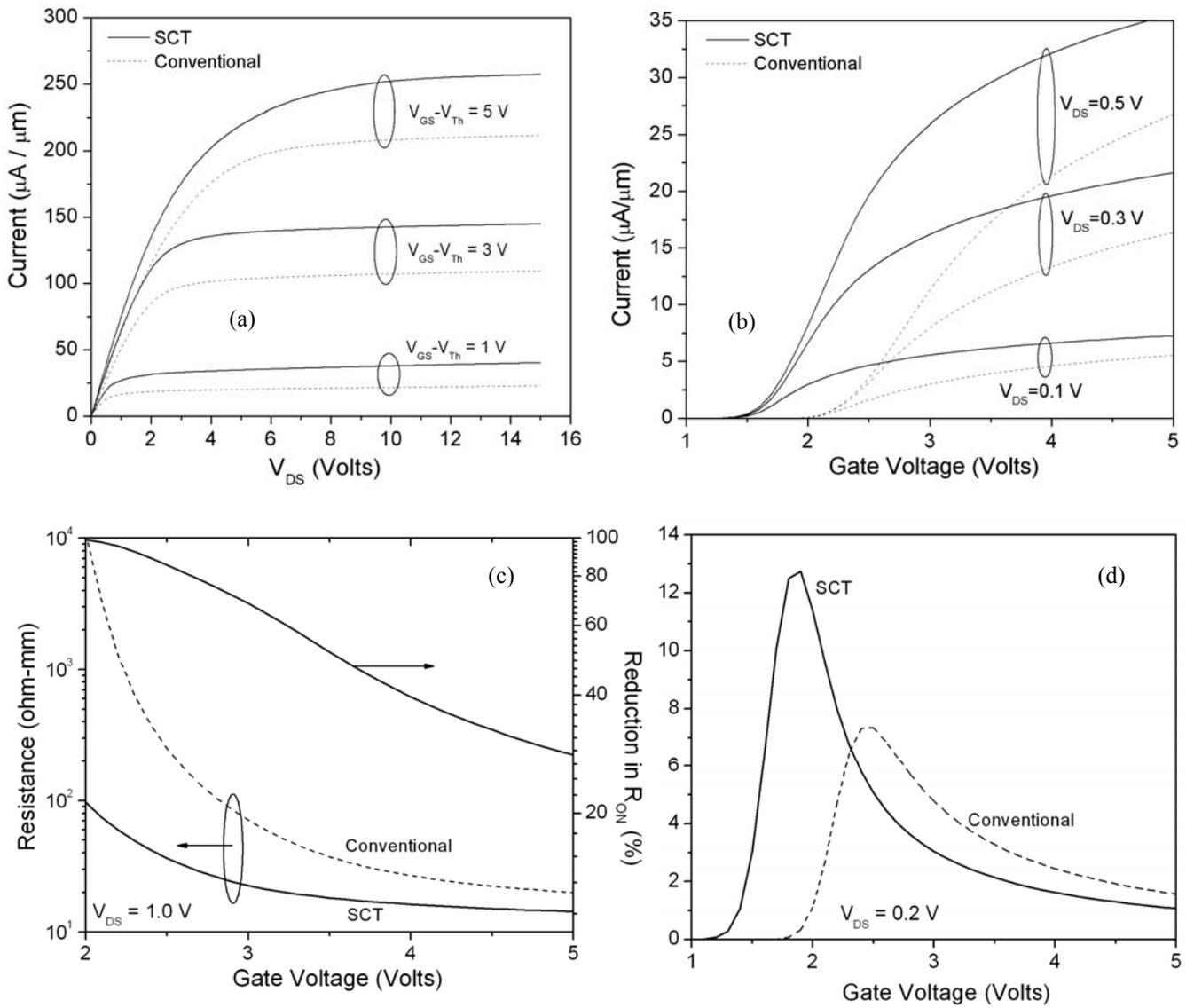

Fig. 5

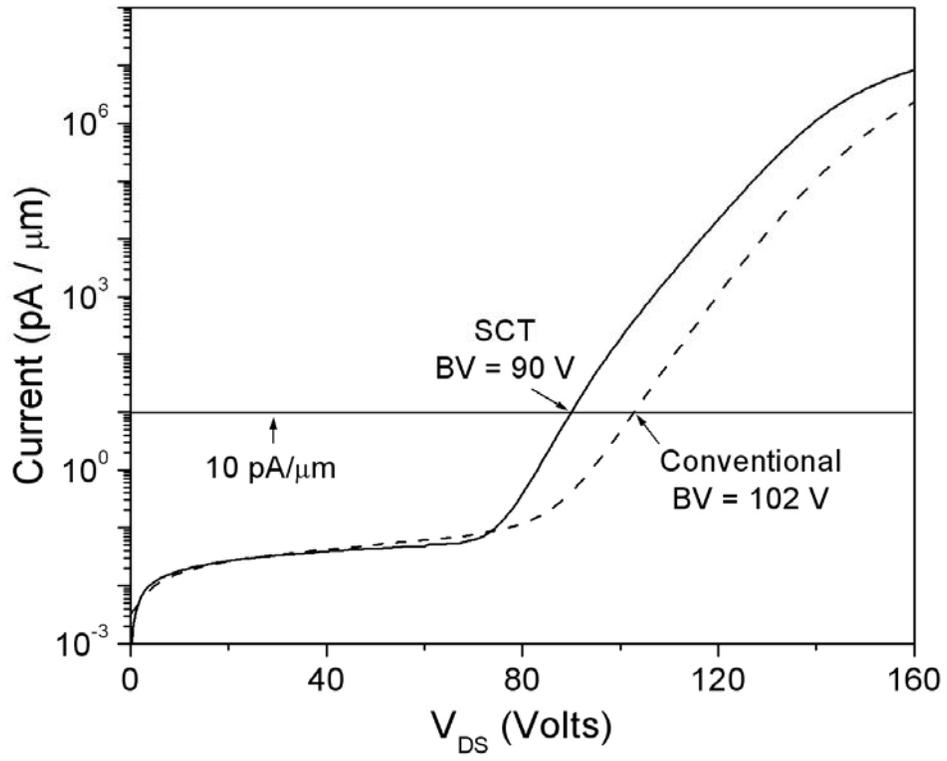

Fig. 6